\documentclass[nolinenumbers, twocolumn]{aastex631}
\usepackage{CJK}

\begin{document}
\newcommand{\LCO}{\affiliation{Las Cumbres Observatory, 6740 Cortona Drive, Suite 102, Goleta, CA 93117-5575, USA}}
\newcommand{\UCSB}{\affiliation{Department of Physics, University of California, Santa Barbara, CA 93106-9530, USA}}
\newcommand{\KITP}{\affiliation{Kavli Institute for Theoretical Physics, University of California, Santa Barbara, 552 University Road, Goleta, CA 93106-4030, USA}}
\newcommand{\UCD}{\affiliation{Department of Physics and Astronomy, University of California, Davis, 1 Shields Avenue, Davis, CA 95616-5270, USA}}
\newcommand{\WIS}{\affiliation{Department of Particle Physics and Astrophysics, Weizmann Institute of Science, 76100 Rehovot, Israel}}
\newcommand{\OKC}{\affiliation{Oskar Klein Centre, Department of Astronomy, Stockholm University, Albanova University Centre, SE-106 91 Stockholm, Sweden}}
\newcommand{\OAPD}{\affiliation{INAF-Osservatorio Astronomico di Padova, Vicolo dell'Osservatorio 5, I-35122 Padova, Italy}}
\newcommand{\Caltech}{\affiliation{Cahill Center for Astronomy and Astrophysics, California Institute of Technology, Mail Code 249-17, Pasadena, CA 91125, USA}}
\newcommand{\GSFC}{\affiliation{Astrophysics Science Division, NASA Goddard Space Flight Center, Mail Code 661, Greenbelt, MD 20771, USA}}
\newcommand{\UMD}{\affiliation{Joint Space-Science Institute, University of Maryland, College Park, MD 20742, USA}}
\newcommand{\UCB}{\affiliation{Department of Astronomy, University of California, Berkeley, CA 94720-3411, USA}}
\newcommand{\TTU}{\affiliation{Department of Physics, Texas Tech University, Box 41051, Lubbock, TX 79409-1051, USA}}
\newcommand{\STScI}{\affiliation{Space Telescope Science Institute, 3700 San Martin Drive, Baltimore, MD 21218-2410, USA}}
\newcommand{\UT}{\affiliation{Department of Astronomy, University of Texas at Austin, 1 University Station C1400, Austin, TX 78712-0259, USA}}
\newcommand{\IoA}{\affiliation{Institute of Astronomy, University of Cambridge, Madingley Road, Cambridge CB3 0HA, UK}}
\newcommand{\QUB}{\affiliation{Astrophysics Research Centre, School of Mathematics and Physics, Queen's University Belfast, Belfast BT7 1NN, UK}}
\newcommand{\IPAC}{\affiliation{Spitzer Science Center, California Institute of Technology, Pasadena, CA 91125, USA}}
\newcommand{\JPL}{\affiliation{Jet Propulsion Laboratory, California Institute of Technology, 4800 Oak Grove Dr, Pasadena, CA 91109, USA}}
\newcommand{\Southampton}{\affiliation{Department of Physics and Astronomy, University of Southampton, Southampton SO17 1BJ, UK}}
\newcommand{\LANL}{\affiliation{Space and Remote Sensing, MS B244, Los Alamos National Laboratory, Los Alamos, NM 87545, USA}}
\newcommand{\Tsinghua}{\affiliation{Physics Department and Tsinghua Center for Astrophysics, Tsinghua University, Beijing, 100084, People's Republic of China}}
\newcommand{\NAOC}{\affiliation{National Astronomical Observatory of China, Chinese Academy of Sciences, Beijing, 100012, People's Republic of China}}
\newcommand{\Itagaki}{\affiliation{Itagaki Astronomical Observatory, Yamagata 990-2492, Japan}}
\newcommand{\Einstein}{\altaffiliation{Einstein Fellow}}
\newcommand{\Hubble}{\altaffiliation{Hubble Fellow}}
\newcommand{\CfA}{\affiliation{Center for Astrophysics \textbar{} Harvard \& Smithsonian, 60 Garden Street, Cambridge, MA 02138-1516, USA}}
\newcommand{\UA}{\affiliation{Steward Observatory, University of Arizona, 933 North Cherry Avenue, Tucson, AZ 85721-0065, USA}}
\newcommand{\MPIA}{\affiliation{Max-Planck-Institut f\"ur Astrophysik, Karl-Schwarzschild-Stra\ss{}e 1, D-85748 Garching, Germany}}
\newcommand{\DSFP}{\altaffiliation{LSSTC Data Science Fellow}}
\newcommand{\HCO}{\affiliation{Harvard College Observatory, 60 Garden Street, Cambridge, MA 02138-1516, USA}}
\newcommand{\Carnegie}{\affiliation{Observatories of the Carnegie Institute for Science, 813 Santa Barbara Street, Pasadena, CA 91101-1232, USA}}
\newcommand{\TAU}{\affiliation{School of Physics and Astronomy, Tel Aviv University, Tel Aviv 69978, Israel}}
\newcommand{\Edinburgh}{\affiliation{Institute for Astronomy, University of Edinburgh, Royal Observatory, Blackford Hill EH9 3HJ, UK}}
\newcommand{\Birmingham}{\affiliation{Birmingham Institute for Gravitational Wave Astronomy and School of Physics and Astronomy, University of Birmingham, Birmingham B15 2TT, UK}}
\newcommand{\Bath}{\affiliation{Department of Physics, University of Bath, Claverton Down, Bath BA2 7AY, UK}}
\newcommand{\CTIO}{\affiliation{Cerro Tololo Inter-American Observatory, National Optical Astronomy Observatory, Casilla 603, La Serena, Chile}}
\newcommand{\Potsdam}{\affiliation{Institut f\"ur Physik und Astronomie, Universit\"at Potsdam, Haus 28, Karl-Liebknecht-Str. 24/25, D-14476 Potsdam-Golm, Germany}}
\newcommand{\INPE}{\affiliation{Instituto Nacional de Pesquisas Espaciais, Avenida dos Astronautas 1758, 12227-010, S\~ao Jos\'e dos Campos -- SP, Brazil}}
\newcommand{\UNC}{\affiliation{Department of Physics and Astronomy, University of North Carolina, 120 East Cameron Avenue, Chapel Hill, NC 27599, USA}}
\newcommand{\Ohio}{\affiliation{Astrophysical Institute, Department of Physics and Astronomy, 251B Clippinger Lab, Ohio University, Athens, OH 45701-2942, USA}}
\newcommand{\AAS}{\affiliation{American Astronomical Society, 1667 K~Street NW, Suite 800, Washington, DC 20006-1681, USA}}
\newcommand{\MMT}{\affiliation{MMT and Steward Observatories, University of Arizona, 933 North Cherry Avenue, Tucson, AZ 85721-0065, USA}}
\newcommand{\Geneva}{\affiliation{ISDC, Department of Astronomy, University of Geneva, Chemin d'\'Ecogia, 16 CH-1290 Versoix, Switzerland}}
\newcommand{\IUCAA}{\affiliation{Inter-University Center for Astronomy and Astrophysics, Post Bag 4, Ganeshkhind, Pune, Maharashtra 411007, India}}
\newcommand{\CMU}{\affiliation{Department of Physics, Carnegie Mellon University, 5000 Forbes Avenue, Pittsburgh, PA 15213-3815, USA}}
\newcommand{\NAOJ}{\affiliation{Division of Science, National Astronomical Observatory of Japan, 2-21-1 Osawa, Mitaka, Tokyo 181-8588, Japan}}
\newcommand{\IfA}{\affiliation{Institute for Astronomy, University of Hawai`i, 2680 Woodlawn Drive, Honolulu, HI 96822-1839, USA}}
\newcommand{\UCSC}{\affiliation{Department of Astronomy and Astrophysics, University of California, Santa Cruz, CA 95064-1077, USA}}
\newcommand{\Purdue}{\affiliation{Department of Physics and Astronomy, Purdue University, 525 Northwestern Avenue, West Lafayette, IN 47907-2036, USA}}
\newcommand{\Princeton}{\affiliation{Department of Astrophysical Sciences, Princeton University, 4 Ivy Lane, Princeton, NJ 08540-7219, USA}}
\newcommand{\Moore}{\affiliation{Gordon and Betty Moore Foundation, 1661 Page Mill Road, Palo Alto, CA 94304-1209, USA}}
\newcommand{\Durham}{\affiliation{Department of Physics, Durham University, South Road, Durham, DH1 3LE, UK}}
\newcommand{\JHU}{\affiliation{William H. Miller III Department of Physics and Astronomy, Johns Hopkins University, Baltimore, MD 21218, USA}}
\newcommand{\ARCO}{\affiliation{Astrophysics Research Center of the Open University (ARCO), Department of Natural Sciences, Ra’anana 4353701, Israel}}
\newcommand{\Toronto}{\affiliation{David A.\ Dunlap Department of Astronomy and Astrophysics, University of Toronto,\\ 50 St.\ George Street, Toronto, Ontario, M5S 3H4 Canada}}
\newcommand{\Duke}{\affiliation{Department of Physics, Duke University, Campus Box 90305, Durham, NC 27708, USA}}
\newcommand{\NCU}{\affiliation{Graduate Institute of Astronomy, National Central University, 300 Jhongda Road, 32001 Jhongli, Taiwan}}
\newcommand{\Columbia}{\affiliation{Department of Physics and Columbia Astrophysics Laboratory, Columbia University, Pupin Hall, New York, NY 10027, USA}}
\newcommand{\Flatiron}{\affiliation{Center for Computational Astrophysics, Flatiron Institute, 162 5th Avenue, New York, NY 10010-5902, USA}}
\newcommand{\CIERA}{\affiliation{Center for Interdisciplinary Exploration and Research in Astrophysics and Department of Physics and Astronomy, \\Northwestern University, 1800 Sherman Avenue, 8th Floor, Evanston, IL 60201, USA}}
\newcommand{\GeminiNorth}{\affiliation{Gemini Observatory, 670 North A`ohoku Place, Hilo, HI 96720-2700, USA}}
\newcommand{\Keck}{\affiliation{W.~M.~Keck Observatory, 65-1120 M\=amalahoa Highway, Kamuela, HI 96743-8431, USA}}
\newcommand{\UW}{\affiliation{Department of Astronomy, University of Washington, 3910 15th Avenue NE, Seattle, WA 98195-0002, USA}}
\newcommand{\DiRAC}{\altaffiliation{DiRAC Fellow}}
\newcommand{\USask}{\affiliation{Department of Physics \& Engineering Physics, University of Saskatchewan, 116 Science Place, Saskatoon, SK S7N 5E2, Canada}}
\newcommand{\Thacher}{\affiliation{Thacher School, 5025 Thacher Road, Ojai, CA 93023-8304, USA}}
\newcommand{\Rutgers}{\affiliation{Department of Physics and Astronomy, Rutgers, the State University of New Jersey,\\136 Frelinghuysen Road, Piscataway, NJ 08854-8019, USA}}
\newcommand{\FSU}{\affiliation{Department of Physics, Florida State University, 77 Chieftan Way, Tallahassee, FL 32306-4350, USA}}
\newcommand{\Melbourne}{\affiliation{School of Physics, The University of Melbourne, Parkville, VIC 3010, Australia}}
\newcommand{\ASTROthreeD}{\affiliation{ARC Centre of Excellence for All Sky Astrophysics in 3 Dimensions (ASTRO 3D)}}
\newcommand{\Stromlo}{\affiliation{Mt.\ Stromlo Observatory, The Research School of Astronomy and Astrophysics, Australian National University, ACT 2601, Australia}}
\newcommand{\NCPAS}{\affiliation{National Centre for the Public Awareness of Science, Australian National University, Canberra, ACT 2611, Australia}}
\newcommand{\TAMU}{\affiliation{Department of Physics and Astronomy, Texas A\&M University, 4242 TAMU, College Station, TX 77843, USA}}
\newcommand{\Mitchell}{\affiliation{George P.\ and Cynthia Woods Mitchell Institute for Fundamental Physics \& Astronomy, College Station, TX 77843, USA}}
\newcommand{\ESO}{\affiliation{European Southern Observatory, Alonso de C\'ordova 3107, Casilla 19, Santiago, Chile}}
\newcommand{\ICE}{\affiliation{Institute of Space Sciences (ICE, CSIC), Campus UAB, Carrer
de Can Magrans, s/n, E-08193 Barcelona, Spain}}
\newcommand{\IEEC}{\affiliation{Institut d'Estudis Espacials de Catalunya, Gran Capit\`a, 2-4, Edifici Nexus, Desp.\ 201, E-08034 Barcelona, Spain}}
\newcommand{\Warwick}{\affiliation{Department of Physics, University of Warwick, Gibbet Hill Road, Coventry CV4 7AL, UK}}
\newcommand{\Macquarie}{\affiliation{School of Mathematical and Physical Sciences, Macquarie University, NSW 2109, Australia}}
\newcommand{\AAARC}{\affiliation{Astronomy, Astrophysics and Astrophotonics Research Centre, Macquarie University, Sydney, NSW 2109, Australia}}
\newcommand{\Capodimonte}{\affiliation{INAF - Capodimonte Astronomical Observatory, Salita Moiariello 16, I-80131 Napoli, Italy}}
\newcommand{\INFNNapoli}{\affiliation{INFN - Napoli, Strada Comunale Cinthia, I-80126 Napoli, Italy}}
\newcommand{\ICRANet}{\affiliation{ICRANet, Piazza della Repubblica 10, I-65122 Pescara, Italy}}
\newcommand{\MSU}{\affiliation{Center for Data Intensive and Time Domain Astronomy, Department of Physics and Astronomy,\\Michigan State University, East Lansing, MI 48824, USA}}
\newcommand{\OSU}{\affiliation{Center for Cosmology and AstroParticle Physics, Ohio State University, 191 West Woodruff Avenue, Columbus, OH 43210-1117, USA}}
\newcommand{\IAIFI}{\affiliation{The NSF AI Institute for Artificial Intelligence and Fundamental Interactions, USA}}
\newcommand{\Warsaw}{\affiliation{Astronomical Observatory, University of Warsaw, Al.\ Ujazdowskie 4, 00-478 Warszawa, Poland}}
\newcommand{\SAAO}{\affiliation{South African Astronomical Observatory, PO Box 9, Observatory 7935, South Africa}}
\newcommand{\UFS}{\affiliation{Department of Physics, University of the Free State, PO Box 339, Bloemfontein 9300, South Africa}}
\newcommand{\UCT}{\affiliation{Department of Astronomy, University of Cape Town, Private Bag X3, Rondebosch 7701, South Africa}}
\newcommand{\Catalyst}{\altaffiliation{LSSTC Catalyst Fellow}}
\newcommand{\UVa}{\affiliation{Department of Astronomy, 530 McCormick Road, Charlottesville, VA 22904-4325, USA}}
\newcommand{\UTA}{\affiliation{Department of Physics, University of Texas at Arlington, Box 19059, Arlington, TX 76019, USA}}
\newcommand{\Konkoly}{\affiliation{Konkoly Observatory, CSFK, MTA Center of Excellence, Konkoly-Thege M. \'ut 15-17, Budapest, 1121, Hungary}}
\newcommand{\ELTE}{\affiliation{ELTE E\"otv\"os Lor\'and University, Institute of Physics and Astronomy, P\'azm\'any P\'eter s\'et\'any 1/A, Budapest, 1117 Hungary}}
\newcommand{\Szeged}{\affiliation{Department of Experimental Physics, University of Szeged, D\'om t\'er 9, Szeged, 6720, Hungary}}
\newcommand{\IAP}{\affiliation{Institut d'Astrophysique de Paris, CNRS--Sorbonne Universit\'e, 98 bis boulevard Arago, F-75014 Paris, France}}
\newcommand{\UCSD}{\affiliation{Department of Astronomy \& Astrophysics, University of California, San Diego, 9500 Gilman Drive, La Jolla, CA 92093-0424, USA}}
\newcommand{\Monash}{\affiliation{School of Physics and Astronomy, Monash University, Clayton, Victoria 3800, Australia}}
\newcommand{\OzGrav}{\affiliation{OzGrav: The ARC Centre of Excellence for Gravitational Wave Discovery, Clayton, Victoria 3800, Australia}}
\newcommand{\Kagoshima}{ \affiliation{Amanogawa Galaxy Astronomy Research Center (AGARC), Graduate School of Science and Engineering, Kagoshima University, 1-21-35 Korimoto, Kagoshima, Kagoshima 890-0065, Japan} }
\newcommand{\ARIES}{ \affiliation{Aryabhatta Research Institute of Observational Sciences, Manora Peak, Nainital 263001, India} }
\newcommand{\MJPRU}{ \affiliation{Department of Applied Physics, Mahatma Jyotiba Phule Rohilkhand University, Bareilly 243006, India} }
\newcommand{\Stockholm}{ \affiliation{Oskar Klein Centre, Department of Astronomy, Stockholm University, AlbaNova, SE-106 91 Stockholm, Sweden} }
\newcommand{\Brera}{ \affiliation{INAF--Brera Astronomical Observatory, Italy} }
\newcommand{\IIA}{ \affiliation{Indian Institute of Astrophysics, Koramangala 2nd Block, Bangalore 560034, India} }
\newcommand{\Hiroshima}{ \affiliation{Hiroshima Astrophysical Science Centre, Hiroshima University, 1-3-1 Kagamiyama, Higashi-Hiroshima, Hiroshima 739-8526, Japan} }
\newcommand{\TDLI}{ \affiliation{Astronomy and Astrophysics Division, Tsung-Dao Lee Institute, Shanghai Jiao Tong University, China} }
\newcommand{\Adler}{\affiliation{Adler Planetarium, 1300 South DuSable Lake Shore Drive, Chicago, IL 60605, USA}}

\title{Early Spectroscopy of the Type~II SN~2024phv Suggests Short-Lived Circumstellar Interaction}

\correspondingauthor{Lucas Aldea}

\author[0009-0009-6114-5899]{Lucas Aldea}
\UCSD
\email{laldea@ucsd.edu}

\author[0000-0002-0832-2974]{Griffin Hosseinzadeh}
\UCSD

\author[0000-0002-4951-8762]{Sergiy S.\ Vasylyev}
\UCSD

\author[0009-0007-7686-5587]{Qiangang Wang}
\UCSD

\author[0000-0002-0744-0047]{Jeniveve Pearson}
\UCSD

\author[0000-0002-7937-6371]{Yize Dong \begin{CJK*}{UTF8}{gbsn}(董一泽)\end{CJK*}\!\!}
\CfA\IAIFI

\author[0000-0003-0123-0062]{Jennifer E.\ Andrews}
\GeminiNorth

\author[0000-0002-4924-444X]{K. Azalee Bostroem}
\IPAC

\author[0000-0002-9454-1742]{Brian Hsu}
\UA

\author[0000-0001-8738-6011]{Saurabh~W.~Jha}
\Rutgers

\author[0000-0002-7352-7845]{Aravind P.\ Ravi}
\UA

\author[0000-0003-4102-380X]{David J.\ Sand}
\UA

\author[0000-0002-4022-1874]{Manisha Shrestha}
\Monash\OzGrav

\author[0000-0001-8073-8731]{Bhagya M.\ Subrayan}
\UA

\author[0000-0002-1895-6639]{Moira Andrews}
\LCO\UCSB

\author[0000-0003-4914-5625]{Joseph R.\ Farah}
\UCB

\author[0000-0003-4253-656X]{D. Andrew Howell}
\LCO\UCSB

\author[0000-0001-5807-7893]{Curtis McCully}
\LCO

\author[0000-0001-9570-0584]{Megan Newsome}
\UT

\author[0000-0003-3656-5268]{Yuan Qi Ni}
\KITP\LCO

\author[0000-0003-0209-9246]{Estefania Padilla-Gonzalez}
\STScI

\author[0000-0003-0794-5982]{Giacomo Terreran}
\Adler

\author[0009-0006-7296-728X]{Kathryn Wynn}
\LCO\UCSB

\author[0000-0002-6703-805X]{Joshua B.\ Haislip}
\UNC

\author[0000-0003-3642-5484]{Vladimir V.\ Kouprianov}
\UNC

\author[0000-0002-5060-3673]{Daniel E.\ Reichart}
\UNC

\begin{abstract}
We present the high-cadence light curve and rapid spectroscopy of the Type~II SN~2024phv, discovered just 16.8 hours after the estimated explosion time.
The early light curve exhibits a rapid 2~mag rise over the first day of observation, and spectra at 1.7 and 2.6 days reveal weak, narrow emission lines, suggesting moderate interaction with circumstellar material during the first few days after explosion. Models of shock-cooling emission provide a reasonable fit to the light curve after day 2, implying minimal interaction at later epochs, and yield a best-fit progenitor radius of $R=390\pm90\,R_{\odot}$. We also compare our earliest spectra with radiation hydrodynamical models of exploding red supergiants to measure circumstellar densities of ${\sim}10^{-12}$~g~cm$^{-3}$ within a few stellar radii of the progenitor, followed by a sharp dropoff in the density profile. This rapidly decreasing profile leads us to conclude that the progenitor of SN~2024phv had an extended atmosphere and/or closely confined circumstellar material, likely from rapidly increasing or eruptive mass loss prior to explosion. In the context of other supernovae discovered at similarly early phases, SN~2024phv points to a diversity in the mass-loss rates of red supergiant stars in their final years.
\end{abstract}


\section{Introduction} \label{sec:intro}

Type II supernovae (SNe II) result from the collapsing cores of hydrogen-rich red supergiant (RSG) progenitors, the final evolutionary phase of stars with zero-age main sequence (ZAMS) masses of $8-30M_{\odot}$. Multiple lines of evidence have revealed various levels of circumstellar material (CSM) around exploding RSGs.
In particular, ``flash'' spectra taken within a few days of explosion show narrow emission lines from the CSM, ionized through interaction with the hot, expanding ejecta \citep{gal-yam_wolf_2014,yaron_confined_2017}.
The fact that these are seen in a significant fraction of SNe~II \citep{khazov_flash_2016,bruch_large_2021,bruch_prevalence_2023} implies that significant mass loss during the final $\sim$1--10 years of the RSG lifespan is common (though see \citealt{kochanek_physics_2019} and \citealt{van_loon_red_2025} for interpretations that do not require enhanced mass loss). Hydrodynamical modeling has also required extended, lower-density material outside of the photosphere to match the fast-rising early light curves of SNe~II, especially in the bluer bands \citep{morozova_unifying_2017,morozova_measuring_2018}. Capturing these short-lived phenomena requires a concerted observational campaign (see \citealt{sand_nearby_2026} for a review).

Programs that specialize in early discovery and rapid follow-up of SNe have greatly increased the sample of SNe~II with early spectroscopy over the past decade \citep{hosseinzadeh_short-lived_2018,hosseinzadeh_weak_2022,andrews_sn_2019,andrews_sn_2024,andrews_asymmetries_2025,soumagnac_sn_2020,bruch_large_2021,bruch_prevalence_2023,hiramatsu_electron-capture_2021,tartaglia_early_2021,jacobson-galan_final_2022,jacobson-galan_final_2024,terreran_early_2022,bostroem_early_2023,bostroem_sn_2023,pearson_circumstellar_2023,smith_high-resolution_2023,shrestha_evidence_2024,shrestha_extended_2024,meza-retamal_circumstellar_2024,zimmerman_complex_2024,lin_sn_2025,dubey_sn_2026,gerard_sn_2026,ransome_eye_2026}.
However, within this sample, there is a wide diversity of strengths of emission features in the early spectra, even controlling for phase, ranging from the sharp Lorentzian profiles in SN~2023ixf \citep[e.g.,][]{smith_high-resolution_2023} to the weak, broad, flat-topped ``ledge'' feature around 4500--4900 \AA{} of SN~2021yja \citep{hosseinzadeh_weak_2022}.
The ledge feature is alternatively treated as broad, blueshifted \ion{He}{2} \citep[e.g.,][]{bullivant_sn_2018,andrews_sn_2019}, a blend of He, C, and N features \citep[e.g.,][]{soumagnac_sn_2020,bruch_large_2021}, or a signature of an extended RSG atmosphere \citep[i.e., not related to CSM;][]{dessart_explosion_2017}. Exploring this diversity is critical to constraining the range of mass-loss rates in the population of RSGs.

In the absence of CSM, the early light curve is powered by shock-cooling emission, the radiation of energy deposited in the ejecta by the initial SN shock. Models of this emission developed and revised by \cite{rabinak_early_2011}, \cite{sapir_uv_2017}, and \cite{morag_shock_2023,morag_shock_2024} can be used to constrain properties of the progenitor, especially the pre-SN radius, which in turn are important for understanding possible mass-loss mechanisms. However, in many (or even most) cases where the rising light curve was sampled densely enough to constrain these models, they are not able to reproduce the rapid rise, presumably contributed by circumstellar interaction \citep[e.g.,][]{hosseinzadeh_short-lived_2018,hosseinzadeh_weak_2022,hosseinzadeh_shock_2023,andrews_sn_2019,dong_supernova_2021,tartaglia_early_2021,pearson_circumstellar_2023,irani_early_2024,jacobson-galan_final_2024,shrestha_extended_2024,gerard_sn_2026}. Investigating when or if these models apply, and how to interpret the results, is thus an important piece of the mass-loss puzzle.

Here we present photometric and spectroscopic observations and analyses of SN\,2024phv, a nearby SN~II discovered within a few hours of explosion by the Distance Less Than 40~Mpc (DLT40) Survey \citep{tartaglia_early_2018}. Section~\ref{sec:obs} presents discovery details and data acquisition. Section~\ref{sec:analysis} covers our photometric and spectroscopic analysis, constraining the properties of the progenitor system and the explosion. Section~\ref{sec:discussion} places our results in the context of previous SNe~II and investigates the validity of the shock-cooling model, given the presence of CSM.

\begin{figure*}
\includegraphics[width=\textwidth]{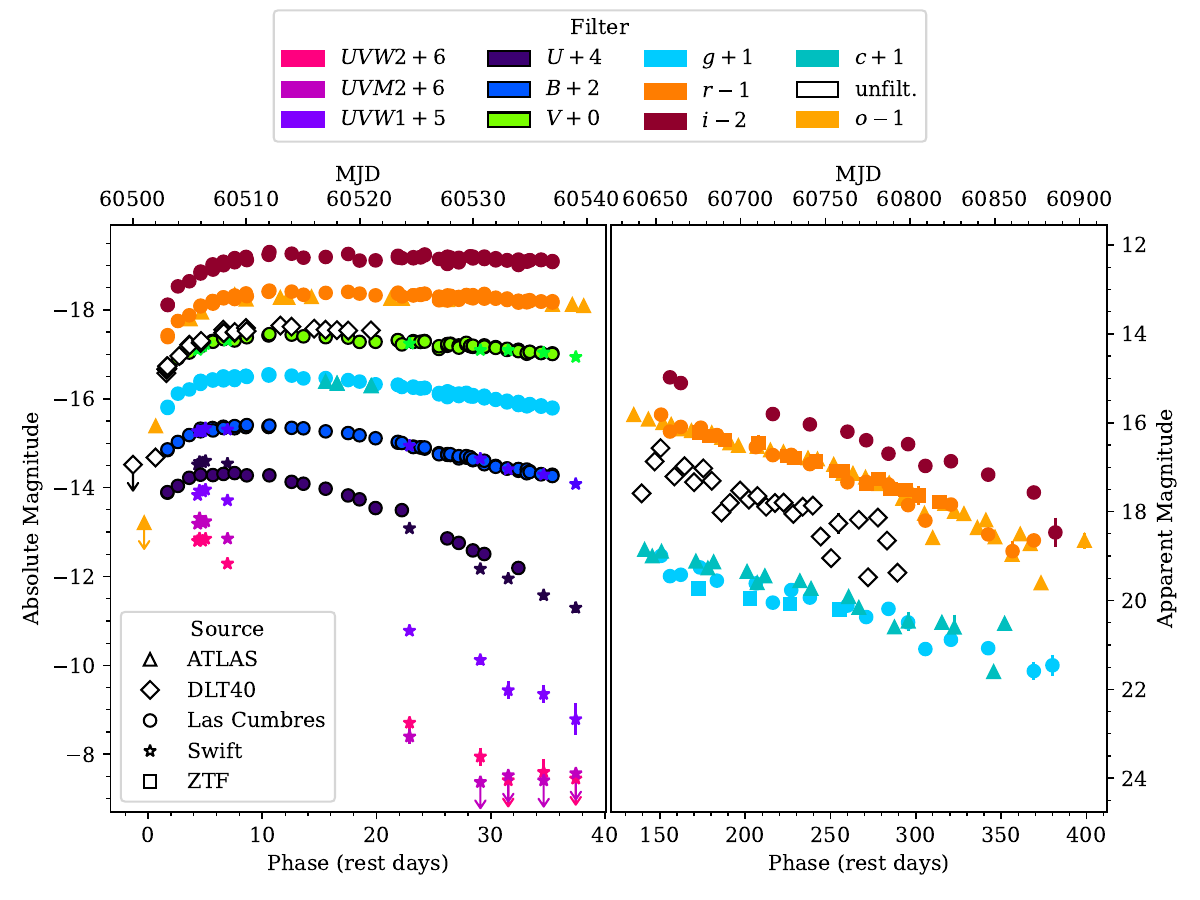}
\caption{The ultraviolet and optical light curve of SN 2024phv DLT40, Las Cumbres Observatory, and Swift, along with publicly available ATLAS and ZTF photometry, in absolute and extinction-corrected apparent magnitudes. The target was sun constrained from days 40--130, during which the fall from the light-curve plateau occurred. We excise this time period from the horizontal axis and bin the late-phase photometry by 5 days for display purposes. Note the deep nondetection the day before discovery and the rapid rise during the following day (see also Figure~\ref{fig:t0}).
\label{fig:lc}}
\end{figure*}

\begin{figure*}
\includegraphics[width=\textwidth]{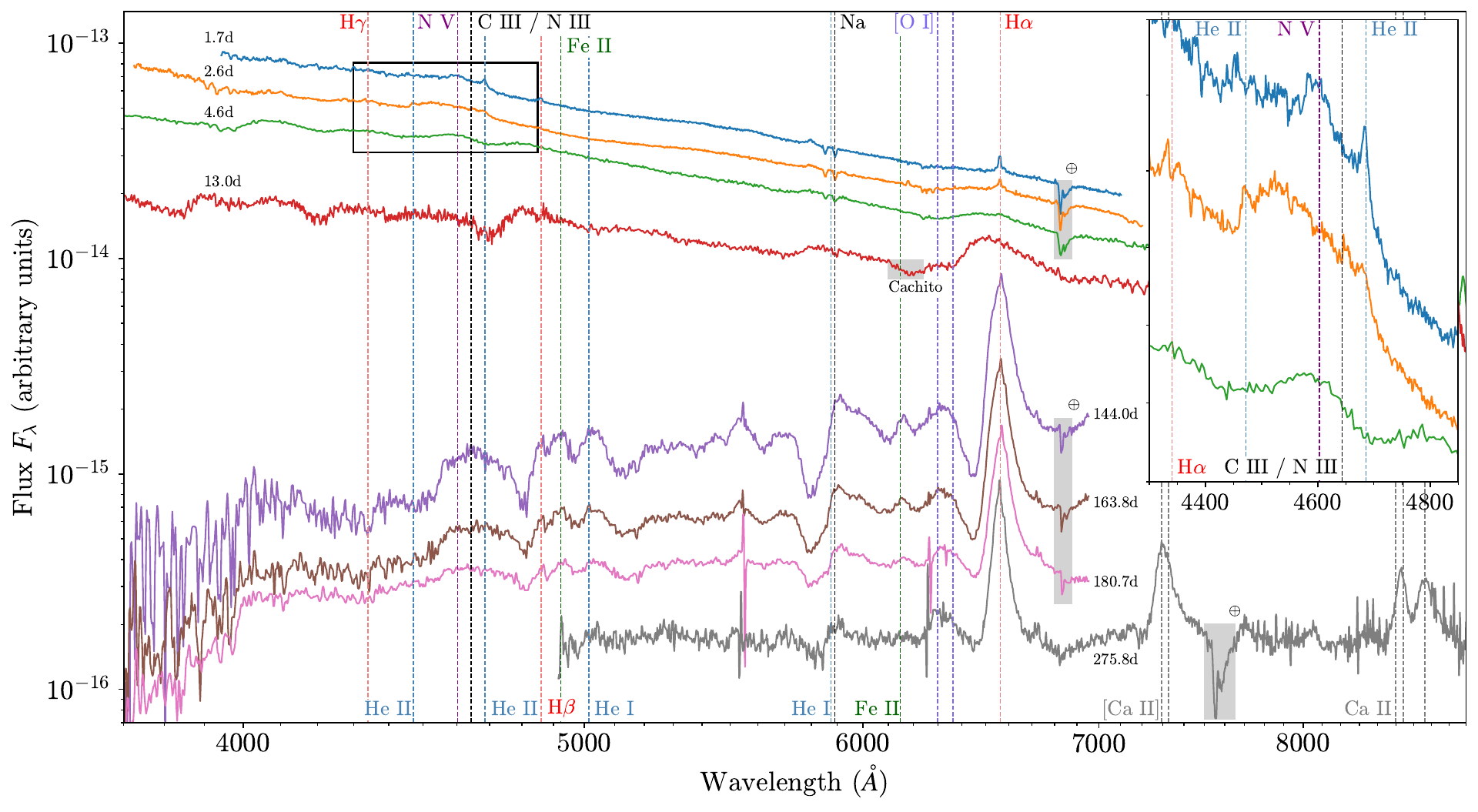}
\caption{All spectra of SN 2024phv logged in Table ~\ref{tab:spec} plotted in log-scale with offsets applied and important elemental lines. The first three spectra are magnified from $\sim4300-4900$\AA{} to highlight the ``ledge" feature and identify possible source lines, including \ion{He}{2}, \ion{C}{3}, \ion{N}{3} and \ion{N}{5}. A second absorption on the blue side of the P Cygni profile of H$\alpha$, dubbed the ``Cachito" feature, at phase 13.0 suggests \ion{Si}{2} presence. Forbidden [\ion{Ca}{2}] and [\ion{O}{1}] emerge in the later spectra, however the SN has not yet reached the nebular regime at phase 275.8.
\label{fig:spec}}
\end{figure*}

\begin{deluxetable*}{lccrrrrrr}[bt!]
\tablecolumns{9}
\tablecaption{Log of Spectroscopic Observations\label{tab:spec}}
\tablehead{
&&& \colhead{Phase} & \multicolumn{5}{c}{Velocity ($10^{3}$kms$^{-1}$)} \\[-10pt]
&&&& \multicolumn{5}{c}{---------------------------------------------} \\[-10pt]
\colhead{MJD} &
\colhead{Telescope} &
\colhead{Instrument} &
\colhead{(d)} &
\colhead{H$\alpha$} &
\colhead{H$\beta$} &
\colhead{H$\gamma$} &
\colhead{H$\delta$} &
\colhead{He I}
}
\startdata
60503.024 & Gemini-S & GMOS & 1.7 & \nodata & \nodata & \nodata & \nodata &\nodata  \\
60503.964 & Gemini-S & GMOS & 2.6 & \nodata & \nodata & \nodata &\nodata  &\nodata  \\
60505.973 & NTT & EFOSC2 & 4.6 & 12.2 & 9.95 & 8.67 & 10.3 & \nodata \\
60514.407 & Faulkes-S & FLOYDS & 13.0 & 11.09 & 9.95 & 9.36 & 8.26 & 8.96 \\
60646.320 & SOAR & Goodman & 145.0 & 5.24 & 3.48 & \nodata & \nodata & \nodata \\
60666.289 & SOAR & Goodman & 163.8 & 5.15 & 3.17 & \nodata &  \nodata&\nodata  \\
60683.245 & SOAR & Goodman & 180.7 & 4.92 & 2.86 &\nodata  & \nodata & \nodata \\
60779.031 & SOAR & Goodman & 275.8 & \nodata& \nodata & \nodata & \nodata & \nodata
\enddata
\tablecomments{Phase is given in rest-frame days after the estimated explosion (MJD 60501.30).}
\end{deluxetable*}
\vspace{-24pt}

\section{Observations} \label{sec:obs}

SN\,2024phv was first observed by DLT40 on 2024 July 10 23:57:02 UTC (MJD 60501.998) at an unfiltered brightness of 17.4 mag \citep{2024TNSTR2371....1V}.
The most constraining nondetection was captured by the Asteroid Terrestrial-impact Last Alert System (ATLAS; \citealt{tonry_atlas_2018a}), providing a limit of $o > 19.22$ mag 1.97 days prior to first detection.
We classified the SN as Type~II the following night, 2024 July 12 00:33:54 (MJD 60503.024), using the Gemini Multi-Object Spectrograph (GMOS) on the Gemini South telescope \citep{hook_gemini_2004,gimeno_on-sky_2016}, highlighting a blue continuum and prominent narrow emission lines of hydrogen and helium \citep{2024TNSCR2403....1A}.

We initiated photometric follow-up of the SN shortly after discovery, capturing \textit{UBVgri} photometry with the 0.4\,m and 1\,m telescopes of Las Cumbres Observatory \citep{brown_cumbres_2013} as part of the Global Supernova Project \citep{howell_global_2017}, as well as continued unfiltered photometry from DLT40, for 34 days after the first detection, at which point the target was sun constrained. We extracted PSF photometry from the Las Cumbres images using \texttt{lcogtsnpipe} \citep{valenti_diversity_2016}. The \textit{UBV} filters are calibrated to Vega magnitudes from the \cite{landolt_ubvri_2009} standard fields observed with the same telescopes on the same nights. The \textit{gri} filters are calibrated to AB magnitudes from the Pan-STARRS1 catalog \citep{chambers_pan-starrs1_2016,flewelling_pan-starrs1_2020}. DLT40 photometry consists of aperture magnitudes calibrated to $r$-band in the AAVSO Photometric All-Sky Survey \citep{henden_aavso_2009}.

We also obtained 10 epochs of photometry from the Ultraviolet/Optical Telescope on the Neil Gehrels Swift Observatory \citep{gehrels_swift_2004}. We measured background-subtracted aperture photometry and applied the updated zero-points of \cite{breeveld_updated_2011} with the 2020 time-dependent sensitivity corrections.

We resumed Las Cumbres and DLT40 photometry 134 days after the first detection as the SN was no longer sun-constrained and continued for 156 days, until it had faded below our detection threshold. For the second season of Las Cumbres photometry, we subtracted reference images from Pan-STARRS1 using PyZOGY \citep{guevel_pyzogy_2021} before measuring photometry. The full light curve is shown in Figure~\ref{fig:lc}.

We obtained an additional GMOS spectrum the night after classification, followed by a spectrum with the ESO Faint Object Spectrograph and Camera 2 (EFOSC2; \citealt{buzzoni_eso_1984}) on the New Technology Telescope (NTT) as part of the Advanced Public ESO Spectroscopic Survey of Transient Objects (ePESSTO+; \citealt{smartt_pessto_2015}),
a spectrum with FLOYDS on Las Cumbres' Faulkes Telescope South \citep{brown_cumbres_2013}, and four spectra with the Goodman High-Throughput Spectrograph \citep{clemens_goodman_2004} on the Southern Astrophysical Research Telescope (SOAR) over the course of the next 9 months. These spectra are logged in Table~\ref{tab:spec} and plotted in Figure~\ref{fig:spec}.

The SN occurred at R.A. = 11\textsuperscript{h}52\textsuperscript{m}23\fs400 and decl. = $-26^\circ54'04\farcs92$, $40''$ from the center of the edge-on spiral galaxy NGC~3936. 
NGC~3936 has a redshift of $z = 0.006715 \pm 0.000010$ \citep{fisher_iras_1995} and a distance modulus of $\mu = 31.42 \pm 0.20$ mag \citep[luminosity distance $d_L = 19.2$~Mpc;][]{tully_cosmicflows-2_2013}.
The Milky Way reddening in the direction of the SN is $E(B-V)_\mathrm{MW} = 0.0696$~mag \citep{schlafly_measuring_2011}.
We calculated the host-galaxy reddening by measuring the equivalent width of the \ion{Na}{1}~D line at the redshift of NGC~3936 and applying the \cite{poznanski_empirical_2012} relation, yielding $E(B-V)_\mathrm{host}=0.091 \pm 0.017$~mag. 
We calculate absolute magnitudes using a \cite{fitzpatrick_correcting_1999} extinction law. 

\section{Analysis} \label{sec:analysis}

\subsection{Early Light Curve}\label{subsec:earlylc}

\begin{figure}[bt!]
\includegraphics[width=\columnwidth]{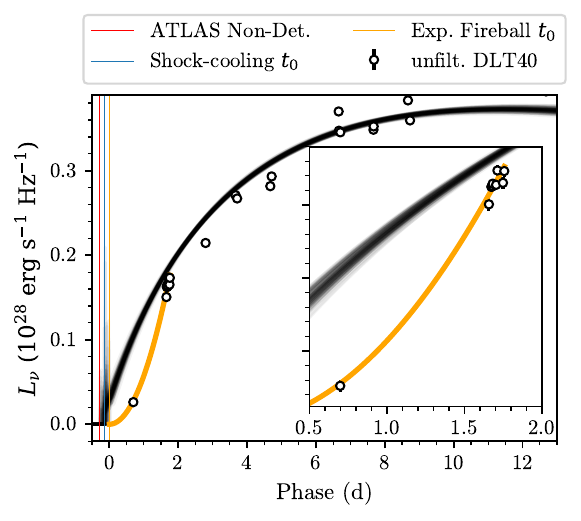}
\caption{Estimating the explosion epoch of SN~2024phv by fitting the first two epochs (8 images) of the DLT40 light curve (circles; error bars are mostly smaller than the markers) with an expanding-fireball model (orange). The shock-cooling model calculated in Section~\ref{subsec:shockcooling} and the most constraining nondetection from ATLAS are shown in black and red, respectively, and the inset shows a magnified view of the first two epochs. While both the shock-cooling model and the expanding-fireball model yield a consistent explosion time estimates (blue and orange vertical lines, respectively), the shock-cooling model greatly overpredicts our first observation and cannot produce the rapid rise over the first 2 days.
\label{fig:t0}}
\end{figure}

The faint absolute magnitude at discovery ($-14$~mag), the nondetection less than a day earlier, and the 2~mag rise over the following day suggest that SN~2024phv was discovered very shortly after explosion. To estimate the explosion epoch ($t_0$), we fit an expanding fireball model, $F \propto (t - t_0)^2$, to the first two epochs (8 images total) of unfiltered DLT40 fluxes ($F$) using weighted least squares. The expanding fireball assumes only homologous expansion of the photosphere and that the shape of the SED does not change much in the optical over 1 day. The fit yields $\mathrm{MJD}_0 = 60501.30 \pm 0.11$, implying that our first observation occurred only $17 \pm 3$~h after explosion. We adopt this as phase = 0 throughout this work. Note that this estimate agrees well with the shock-cooling explosion epoch estimate (MJD $60501.16\pm0.07$) calculated in Section~\ref{subsec:shockcooling}, but importantly, the shock-cooling model cannot reproduce the very rapid rise between the first two epochs of observation (Figure~\ref{fig:t0}).

\begin{figure}
\includegraphics[width=\columnwidth]{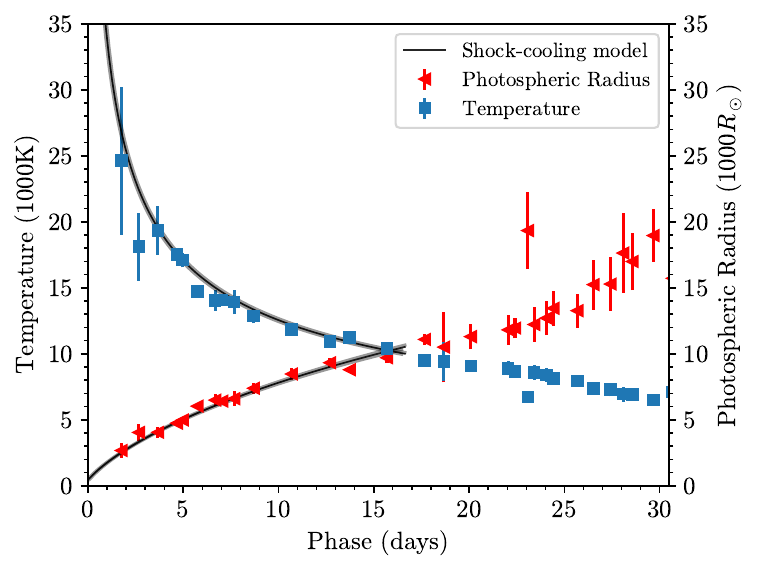}
\caption{The temperature and radius evolution from fitting the Planck function at each epoch of observation (markers) and the evolution implied by the best-fit shock-cooling model (black curves). The spectral energy distribution peaks in the UV shortly after explosion. However, there are no Swift observations for the first 3 epochs, and thus the temperature uncertainty is high.
\label{fig:temprad}}
\end{figure}

We also calculate the bolometric light curve by fitting the Planck function to
the spectral energy distribution (SED) at each epoch of observation. The temperature and photospheric radius were treated as free parameters, estimated via a Markov Chain Monte Carlo (MCMC) routine.
The results are shown in Figure~\ref{fig:temprad}.
Note that the temperature is likely underestimated during the first few days after explosion, as the SED peaked in the UV but Swift observations had not yet begun.
The evolution is typical for SNe of this type, with temperatures ranging from $>$24,000~K to 6,500~K and photospheric radii expanding from 3,000 to 18,000 $R_\odot$ over the first month after explosion. The bolometric light curve is compared to other SNe in Section~\ref{subsec:comparison}.

\subsection{Shock-Cooling}\label{subsec:shockcooling}

\begin{figure*}
\includegraphics[width=\textwidth]{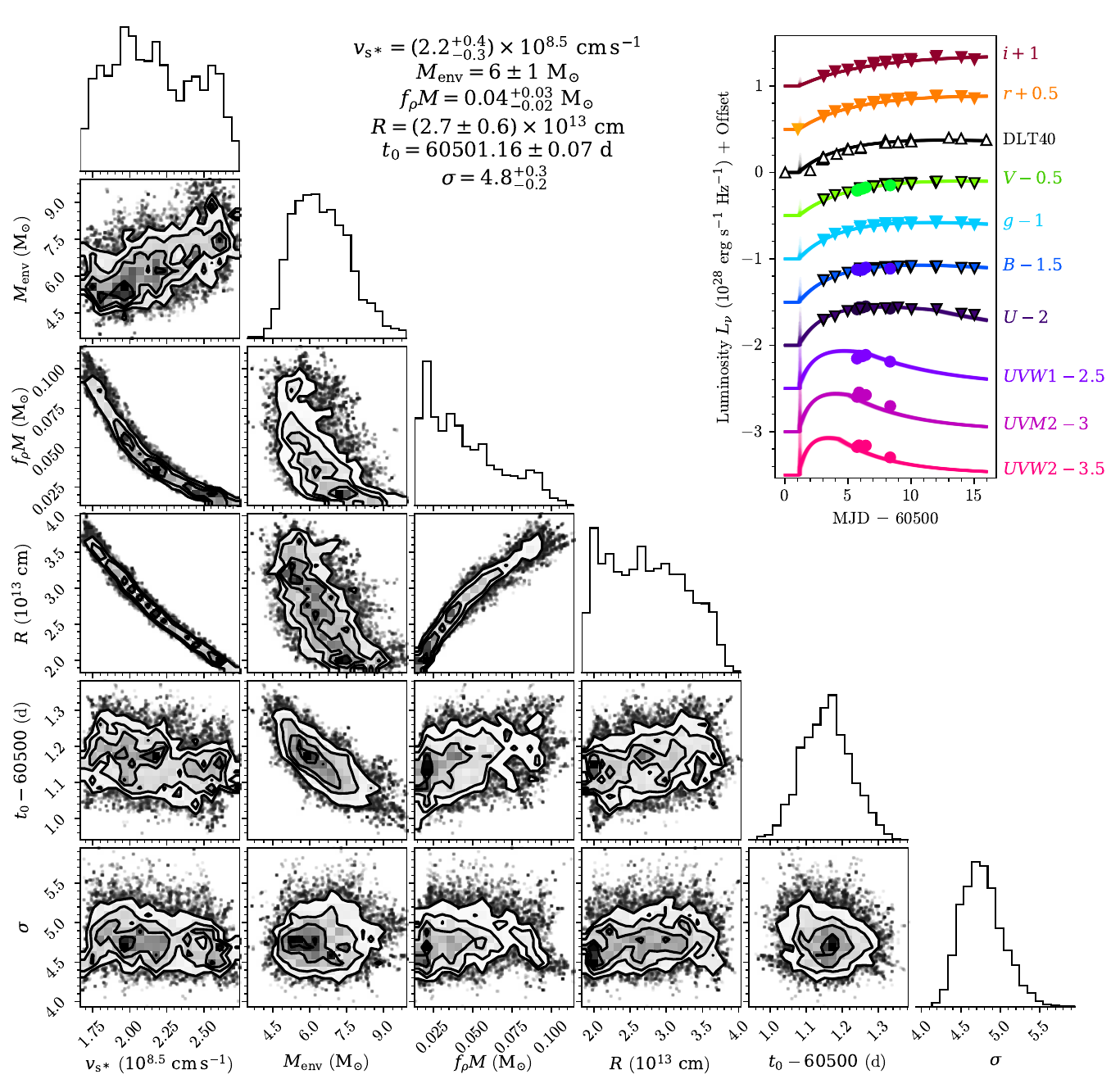}
\caption{Results from fitting the shock-cooling model of \cite{morag_shock_2023} to the first 16 days of light-curve data for all filters, including UV data from Swift and nondetections from ATLAS and DLT40, via an MCMC routine. The corner plot (bottom left) shows the posterior probability distributions for and correlations between each parameter, summarized in Table~\ref{tab:Shock_Cooling_4}. The model provides good fits for all filters, with the exception of the discovery image from DLT40 (top right panel). The parameters and uncertainties reported at the top center represent the 16th, 50th, and 84th percentiles of the posteriors.
\label{fig:shockcooling}}
\end{figure*}

Shock cooling refers to the phase immediately after shock breakout during which the energy deposited by the SN shock is radiated away by the expanding ejecta.
In the absence of circumstellar interaction, this is the dominant power source until the onset of hydrogen recombination when the ejecta reach $\approx$8000~K.
The shock-cooling emission is sensitive to the shock-breakout radius and can therefore be used to estimate the radius of the progenitor star \citep{rabinak_early_2011,sapir_uv_2017,morag_shock_2023,morag_shock_2024}.
SN~2024phv is an interesting test case for the shock-cooling models, as its early spectra suggest weaker circumstellar interaction than in some previous cases (see Section~\ref{sec:spectra}).

The shock-cooling framework models the progenitor star as a polytrope with index $n=\frac{3}{2}$, given a density profile of $\rho_0 = \frac{3f_\rho M}{4\pi R^2}\delta^n$, where $f_\rho$ is a numerical factor of order unity, $M$ is the mass of the ejecta (ignoring possible the remnant), $R$ is the stellar radius, and the fractional depth from the stellar surface is defined as $\delta\equiv\frac{R-r}{R}$. In the shock-cooling framework, $f_\rho M$  only enters the expressions through their product and is therefore treated as a single parameter. The shock velocity is profiled by $v_{sh}=v_{s*}\delta^{-\beta n}$, where $v_{s*}$ is a free parameter, and $\beta=0.191$. The term $M_\mathrm{env}$ is the mass of the outer stellar envelope in the regime $\delta\ll1$. The moment of core-collapse is given by $t_0$.

\begin{deluxetable*}{llccccl}
\caption{Shock Cooling Parameters}
\label{tab:Shock_Cooling_4}
\tablecolumns{7}
\tablehead{
& & \multicolumn{3}{c}{Prior} & Best-fit & \\[-8pt]
& & \multicolumn{3}{c}{------------------------------------} \\[-6pt]
Parameter & Variable & Shape & Min. & Max. & Value & Units}

\startdata
Shock velocity 
& $v_\ast$ 
& Uniform 
& 1.25 
& 2.75 
& $2.2^{+0.4}_{-0.3}$ 
& $10^{8.5}\,\mathrm{cm\,s^{-1}}$ \\
Envelope mass
& $M_{\mathrm{env}}$ 
& Uniform 
& 1 
& 10 
& $6\pm1$ 
& $M_\odot$ \\
Ejecta mass $\times$ numerical factor
& $f_\rho M$ 
& Uniform 
& 0.001 
& 0.15 
& $0.04^{+0.03}_{-0.02}$ 
& $M_\odot$ \\
Progenitor radius 
& $R$ 
& Uniform 
& 1 
& 5 
& $2.7\pm0.6$ 
& $10^{13}\,\mathrm{cm}$ \\
Explosion time 
& $t_0$ 
& Uniform 
& 60500 
& 60503
& $60501.16\pm0.07$ 
& MJD \\
Intrinsic scatter 
& $\sigma$ 
& Log-uniform 
& 0 
& 10 
& $4.8^{+0.3}_{-0.2}$ 
& Dimensionless \\[3pt]
\enddata
\end{deluxetable*}
\vspace{-22pt}

We fit the shock-cooling model of \cite{morag_shock_2023} to the early light curve of SN~2024phv using an MCMC routine implemented in the Light Curve Fitting package \citep{hosseinzadeh_light_2026}, with 12 walkers over an initial 4000 step burn-in followed by an additional 5000 steps to sample the posterior. The model converged well, producing posterior probability distributions and light curves seen in Figure~\ref{fig:shockcooling}, the temperature and radius evolution plotted in Figure~\ref{fig:temprad}, and the best-fit parameters in Table~\ref{tab:Shock_Cooling_4}. In particular, we derive a progenitor radius of $R=390\pm90\,R_{\odot}$, consistent with medium-sized RSGs \citep{levesque_astrophysics_2017}.
Despite the reasonable light-curve fit, radius value, and explosion time, the model overpredicts (by a factor of $\approx$4) the flux in our first image (Figure~\ref{fig:t0}). The model claims to be valid from $\frac{3R}{c} = 65 \pm 8$~min after shock breakout until either the ejecta become optically thin or recombination contributes significantly to the luminosity, which occurs 16.75 days after explosion (MJD 60517.93) for our best-fit parameters.

\subsection{Spectra}\label{sec:spectra}

The first two spectra were taken within a day of each other and within three days of explosion. The early spectra exhibit narrow emission lines in H$\alpha$, H$\beta$, H$\gamma$, and \ion{He}{2}. The narrow lines in \ion{He}{2} and H$\beta$ disappeared in this period, with a ``ledge-shaped'' feature at $\approx$4500--4700 \AA{} becoming more prominent. This feature has been seen previously in, for example, SNe~2017gmr \citep{andrews_sn_2019}, 2018lab \citep{pearson_circumstellar_2023}, 2021gmj \citep{meza-retamal_circumstellar_2024}, 2021yja \citep{hosseinzadeh_weak_2022}, 2022acko \citep{bostroem_sn_2023,lin_sn_2025,teixeira_sn_2026}, 2023axu \citep{shrestha_evidence_2024}, and 2024abfl \citep{gerard_sn_2026}, in some cases interpreted as a sign of (weak) interaction with CSM, but in other cases taken as a sign of an extended RSG atmosphere \citep[e.g.,][]{dessart_explosion_2017}.

In contrast, the spectra captured 4.6 and 13 days after explosion exhibit broad P Cygni profiles in hydrogen and helium, characteristic of SNe~II. The velocities of the components of the ejecta can be calculated from the P~Cygni absorption minima and are recorded in Table ~\ref{tab:spec}. At this phase, the expansion velocity is $\sim$9,000--11,000~km~s$^{-1}$. A second absorption component on the blue side of the H$\alpha$ P~Cygni profile is observed. This feature, dubbed the ``cachito'' by \cite{gutierrez_type_2017}, appears in $60\%$ of observed SNe. \cite{gutierrez_type_2017} suggest that, at phases $<$30 days, the feature is associated with \ion{Si}{2} and is not a sign of high-velocity hydrogen.

Our remaining spectra were taken in the second season of observation, after the fall from the light-curve plateau. The velocities of H$\alpha$ and H$\beta$ fall to 5,000 and 3,000~km~s$^{-1}$, respectively. The forest of absorption lines from iron-group elements is visible at these phases. Our latest spectrum, taken 276 days after explosion, is not fully nebular, but forbidden [\ion{O}{1}] and [\ion{Ca}{2}] are beginning to emerge, along with the \ion{Ca}{2} infrared triplet.

\subsection{Nickel Mass}

At late times, the observed emission is powered predominantly by the radioactive decay of $^{56}$Ni $\to$ $^{56}$Co $\to$ $^{56}$Fe, producing $\gamma$-rays trapped and reradiated by the ejecta, and is characterized by a linear decline in magnitude. The mass of $^{56}$Ni produced in the explosion can be estimated by comparing the pseudo-bolometric luminosity to that of SN~1987A, which has an independent measurement of the $^{56}$Ni production. Our analysis accounts for the incomplete trapping of $\gamma$-rays with the following formula, where $T_{0}$ is the characteristic trapping time scale \citep{clocchiatti_light_1997}:

\begin{eqnarray}
    \frac{L_\mathrm{24phv}}{L_\mathrm{87A}} = \frac{M_\mathrm{Ni,24phv}}{M_\mathrm{Ni,87A}} \times \frac{ 1-\exp\left(-\frac{T_\mathrm{0,24phv}^{2}}{t^{2}}\right)}{ 1-\exp\left(-\frac{T_\mathrm{0,87A}^{2}}{t^{2}}\right)}
\end{eqnarray}

The pseudo-bolometric luminosity of SN~1987A, $L_\mathrm{87A}$, is  calculated using \textit{UBVRI} photometry from \cite{catchpole_spectroscopic_1988}, \cite{suntzeff_sn_1988}, and \cite{whitelock_spectroscopic_1988}. The Ni mass and the trapping time scale of SN~1987A are taken to be $M_\mathrm{Ni,87A}=0.075 M_{\odot}$ and $T_\mathrm{0,87A}=530$~days \citep{jerkstrand_44ti-powered_2011}. The luminosity ratios were fitted using an MCMC routine with a 1500 step burn-in and 500 step posterior sample and yielded the following parameters:
\[M_\mathrm{Ni, 24phv} =0.0232^{+0.0018}_{-0.0015}M_{\odot}\]
\[T_\mathrm{0, 24phv} =350^{+40}_{-30}\mathrm{d} \]
\cite{valenti_diversity_2016} present a distribution of nickel masses in SNe~II ranging from $0.0014\ M_\odot$ to $0.077\ M_\odot$. \cite{muller_nickel_2017} present another sample with a skewed Gaussian distribution between $0.005\ M_\odot$ and $0.280\ M_\odot$, with a mean of $0.046M_\odot$ and a median of $0.031M_\odot$. Compared to these samples, SN~2024phv synthesized a slightly below-average but not unusually low mass of $^{56}$Ni.

\subsection{Pre-explosion Activity Search}
Motivated by the signatures of confined CSM discussed above, we search archival imagery from time-domain surveys for precursor emission---any detection at the position of the SN prior to explosion. We obtained forced photometry at the position of SN~2024phv from the ATLAS forced photometry server \citep{shingles_release_2021} and the ZTF Forced Photometry Service \citep[ZFPS;][]{masci_new_2023}. We stacked the measurements in 15-day bins following \citet{Young2022} and \citet{masci_new_2023}, adopting a $3\sigma$ detection threshold and $5\sigma$ upper limits \citep{Masci2011ComputingFU}. No precursor emission was detected down to approximately $-11$ mag in the available observations spanning the 3,000 days before explosion. If there were precursor activities for SN~2024phv, they would have to be faint and only last for a short duration.

\section{Discussion} \label{sec:discussion}\label{subsec:comparison}

\begin{figure*}
\includegraphics[width=\textwidth]{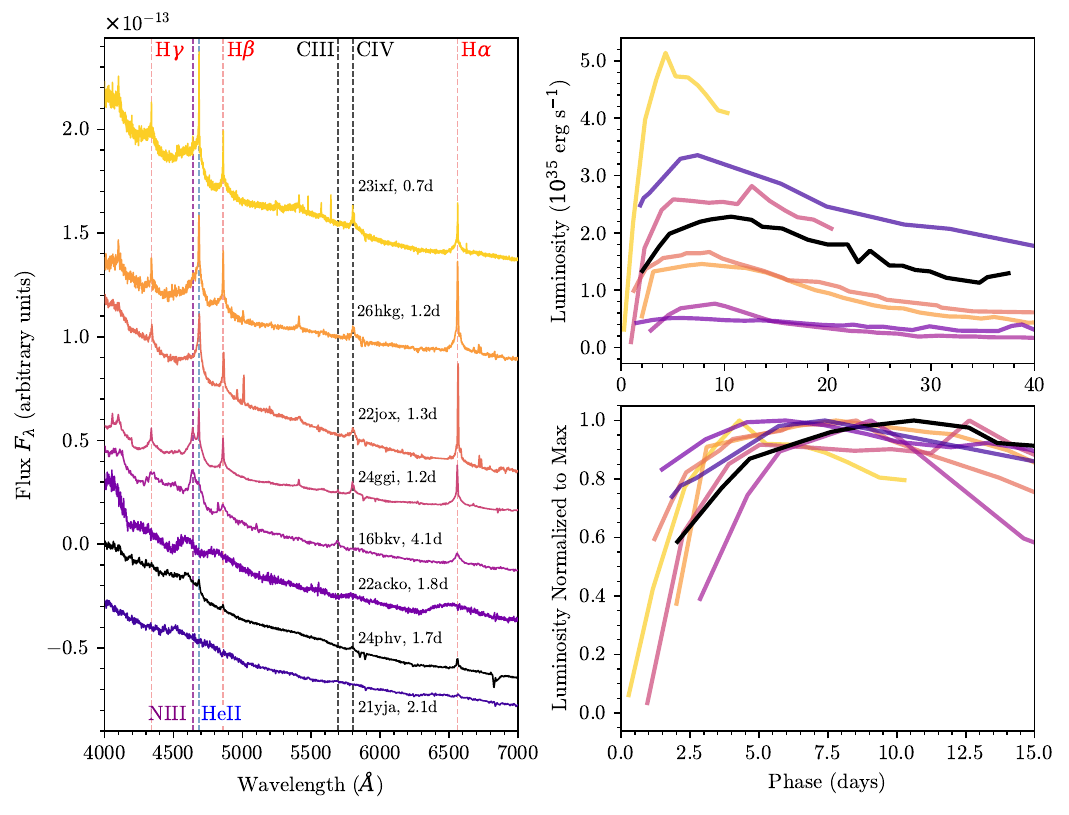}
\caption{Comparison of SN~2024phv to other SNe~II observed within days of explosion: SN 2023ixf, SN 2024hkg, SN 2022jox, SN 2024ggi, SN 2016bkv, SN 2022acko, SN 2024phv, and SN 2021yja. The left panel shows the diversity in their early spectra, sorted and colored by strength of the narrow emission features, showing differences in the level of CSM interaction. The right panels show their pseudobolometric light curves, with the same color scheme as the left panel. The normalized light curves (bottom) reveal a quicker rise to plateau for events with higher CSM interaction, while the unnormalized light curves (top) demonstrate that this effect does not correlate with the peak luminosity. It is unclear from the spectra if the ``ledge'' feature is produced by low to moderate CSM interaction or by an extended progenitor envelope.
\label{fig:lccomp}}
\end{figure*}

Several lines of evidence lead us to the conclusion that SN~2024phv had short-lived, moderate circumstellar interaction, i.e., that its progenitor had an extended stellar atmosphere and/or dense CSM confined near the progenitor at the time of explosion. First, its early spectra showed only weak narrow emission features of H and He. If we compare its early spectra to those of other SNe~II discovered approximately as early---SN 2016bkv \citep{hosseinzadeh_short-lived_2018}, SN~2021yja \citep{hosseinzadeh_weak_2022}, SN~2022jox \citep{andrews_sn_2024}, SN 2022acko \citep{bostroem_sn_2023,teixeira_sn_2026}, SN~2023ixf \citep{bostroem_early_2023,jacobson-galan_sn_2023}, SN~2024ggi \citep{shrestha_extended_2024}, and SN~2026hkg (S.~V.~Vasylyev et al.\ 2026, in preparation)---we see what appears to be a continuum of flash-ionized feature strength, with SN~2024phv near the bottom (Figure~\ref{fig:lccomp}, left). A ``ledge'' feature is present blueward of the \ion{He}{2} line in SN~2024phv and SN~2021yja, also on the low-interaction end of this continuum. When displayed this way, Figure~\ref{fig:lccomp} suggests that the ledge does not correlate with signatures of CSM. \cite{dessart_explosion_2017} concur, suggesting that it may instead indicate an extended RSG atmosphere.

Another signature of circumstellar interaction is known to be a rapid light-curve rise \citep[e.g.,][]{morozova_unifying_2017,morozova_measuring_2018}, so we compare the early pseudobolometric light curve of SN~2024phv to those of the same sample of SNe~II. If we normalize the light curves to their peak value (Fig.~\ref{fig:lccomp}, bottom right), we see that, indeed, the SNe with stronger narrow emission lines tend to rise faster to peak, whereas SN~2024phv peaks the latest, after more than a week. Notably, the strength of the interaction features does not correlate perfectly with the peak luminosity; events with stronger interaction signatures are both brighter and fainter than SN~2024phv (Figure~\ref{fig:lccomp}, top right). As we discuss in Section~\ref{subsec:earlylc}, the overall slow rise to peak is not smooth; the SN is already above half its peak luminosity by day 2, before which we have no reliable estimate of the bolometric luminosity. We interpret this as a sign that the extended atmosphere or dense CSM is tightly confined near the progenitor.

\begin{figure}
\includegraphics[width=\columnwidth]{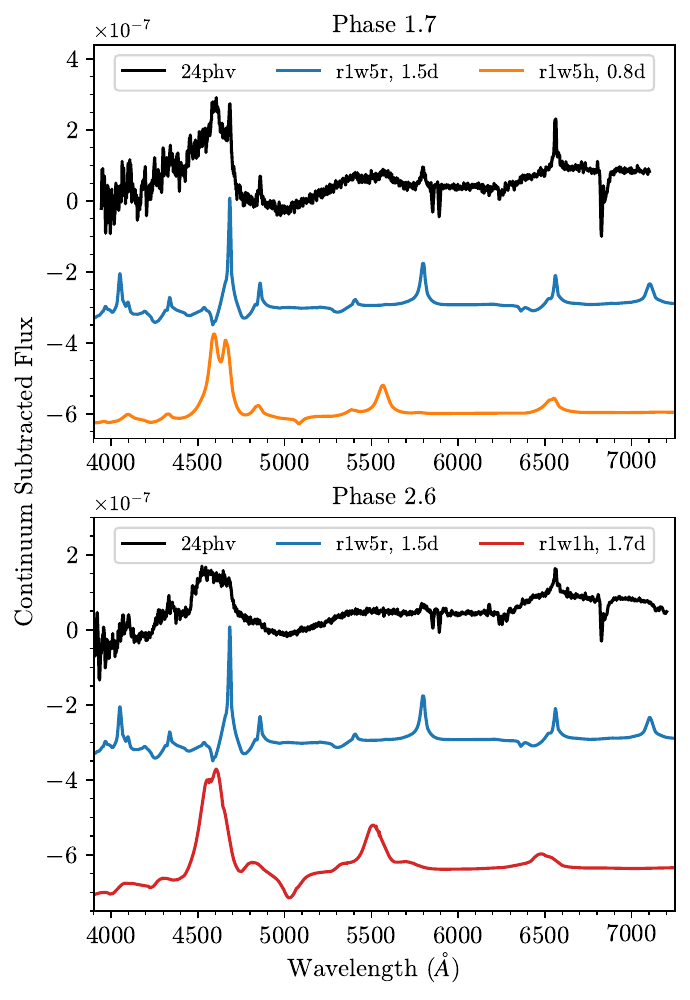}
\caption{The early spectra of SN 2024phv at 1.7~d (top) and 2.6~d (bottom) after explosion, compared with similar radiation hydrodynamical models of RSG explosions from \cite{dessart_explosion_2017}. The narrow emission features of r1w5r at 1.0~d remain present in both spectra, decreasing in intensity by day 2.6. The more ``ledge-shaped'' features of r1w5h at 0.5~d resemble our day-1.7 spectrum, whereas these have flattened to resemble r1w1h at 0.7~d by day 2.6. As we discuss in Section~\ref{sec:discussion}, these all imply a density of ${\sim}10^{-12}$~g~cm$^{-3}$ at a radius of a few times the progenitor radius (${\sim}2\times10^{14}$~cm), followed by a sharp dropoff in the density profile before the next spectrum was taken. This suggests that the progenitor experienced a rapidly increasing mass-loss rate or eruptive mass loss during its last years, or that it developed an extended envelope prior to explosion.
\label{fig:modelspec}}
\end{figure}

\cite{dessart_explosion_2017} present a collection of radiation hydrodynamical models of an RSG progenitor ($R_\star=501R_\odot$) exploding into CSM with various density profiles. In Figure~\ref{fig:modelspec}, we see that SN~2024phv at day 1.7 exhibits characteristics of the r1w5r model at 1.0~d and the r1w5h model at 0.83~d. By day 2.6, the spectrum is better represented by r1w5r at 1.5~d (though with weaker lines) and r1w1h at 1.67~d. Rather than quoting the (constant) mass-loss rates corresponding to these models, we treat these matches as an indication of the CSM density at two different radii. All these models have a photospheric radius of $(1{-}2)\times10^{14}$~cm at 0.5--1.0~d \citep[Fig.~A.1]{dessart_explosion_2017}, just a few times the stellar radius, and, in the case of r1w1h, within the extended atmosphere, probing densities of ${\sim}10^{-12}$~g~cm$^{-3}$ \citep[Fig.~1]{dessart_explosion_2017}. Using a fiducial velocity of 10,000~km~s$^{-1}$ (cf.\ Table~\ref{fig:spec}), the ejecta of SN~2024phv should also have reached a radius of $2\times10^{14}$~cm when our first spectra were taken. The fact that the lines are gone by day 4.6 suggests a very steep density profile around the progenitor of SN~2024phv, corresponding to a rapid increase in mass-loss rate, or indeed eruptive mass loss, during the last years of its life. However, we do not see any precursor emission that could explain this CSM. Alternatively, the higher density in our first spectra could be the result of an extended RSG atmosphere, surrounded by very low density material.

The fit of the shock-cooling model in Section~\ref{subsec:shockcooling} reinforces this argument, as the model does not perfectly fit the first few days due to the substantial interaction. With the sharply declining CSM density profile, the model fit is good from day 2 through the end if its validity range. The model yields an explosion time that is consistent with a simple expanding fireball model and a reasonable progenitor radius for an RSG, $R=390\pm90\,R_{\odot}$. In comparison, an earlier version of the shock-cooling model from \cite{sapir_uv_2017} also fit SN~2021yja fairly well, but yielded an unreasonably large radius \citep{hosseinzadeh_weak_2022}. Improvements in the model SED by \cite{morag_shock_2023} may have made these radius measurements more reliable by improving the early model behavior. However, the model predicts an envelope mass of $M_\mathrm{env}=6M_{\odot}$ and a (scaled) ejecta mass of $f_{\rho}M = 0.04M_{\odot}$, which is not physically possible if $f \approx 1$. A similar issue was presented when fitting  the light curve of SN~2022acko, where the $f_{\rho}M$ parameter was constrained to nonphysically small values \citep{teixeira_sn_2026}. 

\begin{figure}
\includegraphics[width=\columnwidth]{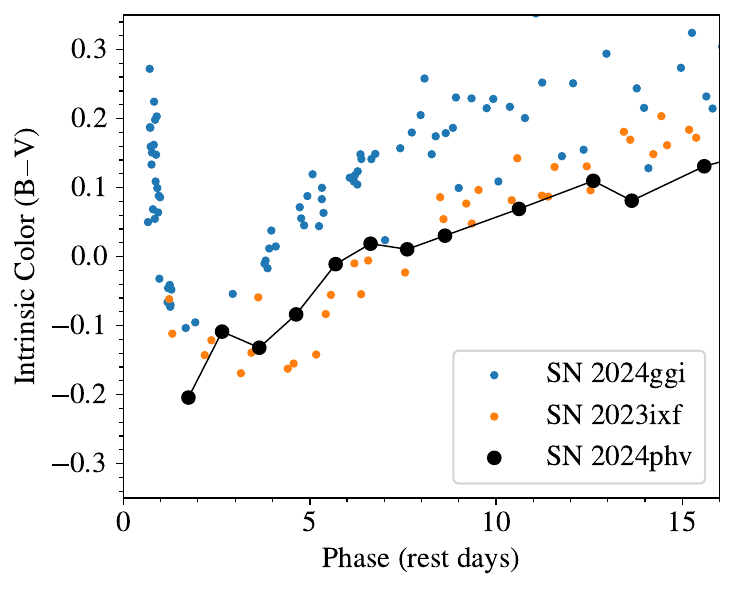}
\caption{The $B-V$ color evolution of SN 2024phv compared to SN 2024ggi and SN 2023ixf from \cite{shrestha_evidence_2024}. The latter two SNe have high CSM interaction and an initial blueward evolution not observed in SN 2024phv. This suggests that, if SN~2024phv did interact strongly with CSM, such interaction ended within $\sim$2 days of explosion.
\label{fig:color}}
\end{figure}

Lastly, we compared the extinction-corrected $B-V$ color curves of SN~2024phv to two other SNe~II discovered exceptionally early, SNe~2023ixf and 2024ggi \citep{shrestha_extended_2024}, in Figure~\ref{fig:color}. There is no sign of the initial blueward evolution seen in these events \citep[and also in][]{andrews_asymmetries_2025}, thought to indicate extended shock breakout through the CSM \citep{shrestha_extended_2024}. If such an effect happened in SN~2024phv, it must have been before we had multiband photometry, $\lesssim$2 days after explosion, again implicating very compact CSM.

Taken as a whole, our analysis paints a coherent picture: SN~2024phv is the explosion of a typical RSG interacting with confined CSM for $\lesssim$2 days before continuing to expand into a relatively clean circumstellar environment. The number of events with comparable early data sets remains small but is increasing rapidly. As we fill in the continuum we observe in Figure~\ref{fig:lccomp}, we can begin to understand the differences in mass-loss rates in the full population of RSGs, and thus constrain the mechanism(s) for such mass loss.

\section{Summary and Conclusions} \label{sec:conclusion}

We have presented multi-band photometric and spectroscopic observations of the Type II SN 2024phv discovered 16.8 hours after explosion.
This discovery was followed by a rapid $\sim$2~mag brightening over the course of the next day.
Its spectral series shows relatively weak and fading narrow emission lines during the first 3 days after explosion, indicating moderate but short-lived circumstellar interaction.

The shock-cooling model of \cite{morag_shock_2023} provides a reasonable fit to the light curve, apart from the initial fast rise, and yields a progenitor radius of $R=390\pm90\,R_{\odot}$, typical of a RSG progenitor. Late-phase photometry constrains the $^{56}$Ni to be $M_\mathrm{Ni} =0.0232^{+0.0018}_{-0.0015}M_{\odot}$, which is also fairly typical for a SN~II.
Compared to other SNe~II discovered similarly early, we find that SN~2024phv exhibits a slower rise to plateau and weaker narrow emission lines in the early spectra, which we interpret as signatures of shorter-lived circumstellar interaction. SN 2024phv has no sign of the early redward color evolution seen in other events discovered within a day of explosion, thought to indicate shock breakout through extended CSM. 

The spectral models of \cite{dessart_explosion_2017} suggest a circumstellar density of ${\sim}2\times10^{-14}$~g~cm$^{-3}$ within a few stellar radii of the progenitor of SN~2024phv. The absence of narrow emission features by day 4.6 implies a sharp dropoff in the density profile beyond that. Thus, we argue that the progenitor likely had an extended stellar envelope or relatively compact CSM confined near the RSG at the time of explosion. The broader ``ledge'' feature seen in the early spectra may be attributable to an extended envelope. This stellar and circumstellar configuration suggests that the progenitor RSG lost only a moderate amount of mass in the years before explosion. Such CSM would be reminiscent of the elevated pre-explosion luminosity of SN~2020tlf \citep{jacobson-galan_final_2022}, but we do not see any such precursor emission associated with SN~2024phv.

The literature now contains a continuum of CSM levels in SNe~II, with SN~2024phv at the low to moderate end. To explore the population further, we must continue to push for early discoveries and rapid follow-up observations of a larger sample of events. By increasing the horizon to which we can discover SNe within a day of explosion, the Vera C.\ Rubin Observatory, in combination with other facilities that can supplement its cadence and conduct spectroscopic follow-up, will accelerate the growth of such a sample.

\begin{acknowledgments}

We thank the ePESSTO+ Collaboration, particularly Santiago Gonz\'alez Gait\'an, Jo\~ao Duarte, and Claudia Guti\'errez, for obtaining and reducing the NTT spectrum.
Based on observations collected at the European Organisation for
Astronomical Research in the Southern Hemisphere under ESO programme 1103.D-0328.

Based on observations obtained at the international Gemini Observatory, a program of NSF NOIRLab, which is managed by the Association of Universities for Research in Astronomy (AURA) under a cooperative agreement with the U.S. National Science Foundation on behalf of the Gemini Observatory partnership: the U.S. National Science Foundation (United States), National Research Council (Canada), Agencia Nacional de Investigaci\'{o}n y Desarrollo (Chile), Ministerio de Ciencia, Tecnolog\'{i}a e Innovaci\'{o}n (Argentina), Minist\'{e}rio da Ci\^{e}ncia, Tecnologia, Inova\c{c}\~{o}es e Comunica\c{c}\~{o}es (Brazil), and Korea Astronomy and Space Science Institute (Republic of Korea).

This work makes use of observations from the Las Cumbres Observatory global telescope network.

Based in part on observations obtained at the Southern Astrophysical Research (SOAR) telescope, which is a joint project of the Minist\'{e}rio da Ci\^{e}ncia, Tecnologia e Inova\c{c}\~{o}es (MCTI/LNA) do Brasil, the U.S. National Science Foundation NOIRLab, the University of North Carolina at Chapel Hill (UNC), and Michigan State University (MSU).

Based in part on observations obtained through the Astronomical Event Observatory Network (AEON), a joint endeavor of the Las Cumbres Observatory and of NSF NOIRLab, which is managed by the Association of Universities for Research in Astronomy (AURA) under a cooperative agreement with the U.S. National Science Foundation.

\end{acknowledgments}

\facilities{ADS, ATLAS, CTIO:PROMPT, FTS (FLOYDS), Gemini:South (GMOS), LCOGT (QHY600, Sinistro), NED, NTT (EFOSC2), PO:1.2m (ZTFCam), SOAR (Goodman), Swift (UVOT), TNS}

\software{Astropy \citep{astropy_collaboration_astropy_2018,astropy_collaboration_astropy_2013,astropy_collaboration_astropy_2022}, BANZAI \citep{mccully_real-time_2018}, \texttt{corner} \citep{foreman-mackey_corner.py_2016}, \texttt{emcee} \citep{foreman-mackey_emcee_2013}, FLOYDS pipeline \citep{valenti_first_2014}, HEAsoft \citep{nasa_heasarc_heasoft_2014}, IPython \citep{perez_ipython_2007}, IRAF \citep{national_optical_astronomy_observatories_iraf_1999}, Jupyter \citep{kluyver_jupyter_2016}, \texttt{lcogtsnpipe} \citep{valenti_diversity_2016}, Light Curve Fitting \citep{hosseinzadeh_light_2026}, Matplotlib \citep{hunter_matplotlib_2007}, NumPy \citep{harris_array_2020}, PyRAF \citep{science_software_branch_at_stsci_pyraf_2012}}

\bibliography{zotero_abbrev}

\end{document}